\documentclass[]{spie}  

\usepackage{amsmath,amsfonts,amssymb}
\usepackage{graphicx}
\usepackage[colorlinks=true, allcolors=blue]{hyperref}

\title{Laboratory testing of the Ingot WFS}

\author[a,c]{Simone Di Filippo}
\author[a,c]{Davide Greggio}
\author[a,c]{Maria Bergomi}
\author[a,c]{Kalyan Kumar Radhakrishnan Santhakumari}
\author[c,d]{Elisa Portaluri}
\author[a,c]{Carmelo Arcidiacono}
\author[a,c]{Valentina Viotto}
\author[a,b,c]{Roberto Ragazzoni}
\author[a,c]{Marco Dima}
\author[a,c]{Luca Marafatto}
\author[a,c]{Jacopo Farinato}
\author[a,c]{Demetrio Magrin}
\affil[a]{INAF -  Osservatorio Astronomico di Padova, Vicolo dell'Osservatorio 5, I-35122,Padova, Italy}
\affil[b]{Dipartimento di Fisica e Astronomia, Università degli Studi di Padova, Vicolo dell’Osservatorio 3, I-35122 Padova, Italy}
\affil[c]{ADONI, Laboratorio Nazionale di Ottica Adattiva, Italy}
\affil[d]{INAF - Osservatorio Astronomico d'Abruzzo, Via Mentore Maggini, I-64100 Teramo, Italy}

\authorinfo{Further author information: simone.difilippo@inaf.it}

\begin{document} 
\maketitle

\begin{abstract}
The ingot WFS is a new kind of wavefront sensor specifically designed to deal with the elongation of LGS reference sources on ELT-class telescopes. Like the pyramid, it belongs to the family of pupil plane wavefront sensors and can be considered as a generalization of the pyramid WFS for extended, three-dimensional elongated sources. The current design uses a simple, reflective roof-shaped prism to split the light into three pupils that are used to retrieve the wavefront shape.  A test-bench has been realized at the INAF-Padova laboratories to test the alignment and functioning of the ingot. The bench is equipped with a deformable lens, conjugated to the pupil plane, able to apply low-order aberrations and with a hexapod for the precise alignment of the ingot prism. In this work we present a robust and fully automated Python-code alignment procedure, which is able, by using the optical feedback from the I-WFS, to adjust its 6-degrees of freedom. Moreover, we report on the tests conducted with the deformable lens to characterize the ingot WFS response to low-order aberrations in terms of sensitivity and linearity. The results are used as a comparison for simulations to validate the ray-tracing modeling approach with the future goal of optimizing the procedure adopted for signal calculation and phase retrieval.  
\end{abstract}


\section{INTRODUCTION}
Despite their name, Sodium Laser Guide Stars (LGSs)\cite{Thompson_Gardner_LGS}, cannot be fully considered real point-like source. They are placed at a finite distance, and if fired from the side of a large telescope, such as the Extremely Large Telescope (ELT)\cite{E-ELT}, their characteristics change dramatically depending on the position on the entrance pupil from where the source is being observed. This effect is due to the intrinsic nature of the Sodium layer, which is located at about 90 km in the atmosphere and extended for about 10-20 km\cite{butler_d_j_measuring_2003}. In addition, this layer has not a fixed thickness, but a peculiar vertical density distribution, that varies both spatially and temporally\cite{pfrommer_high_2014}. The Ingot Wavefront Sensor (I-WFS), presented for the first time by Ragazzoni\cite{ingot0}, aims to cope with these intrinsic characteristics of the LGS, and in particular with the typical 3-dimensional elongation. That elongation produces a non-negligible effect on many of the classical wavefront sensors, like the Shack-Hartmann wavefront sensor (SH-WFS), for which the spots are elongated in the opposite direction of the side of the laser launcher position. \\
In this work, we present the current status of the laboratory testing of the I-WFS, in particular the design of the test bench developed in the INAF - Osservatorio Astronomico di Padova (Italy) laboratory. In this framework, we developed a setup that aims to reproduce as close as possible, the characteristics of the ELT. We artificially reproduced the LGS source to investigate the behavior of the I-WFS in an open loop scenario, developing a fully automatic python procedure to align the I-WFS to the optical axis and so to the LGS movements in the sky. Lastly, having equipped the bench with a deformable lens to be used as an aberrator, we defined a calibration/reconstruction procedure to investigate the response of the I-WFS when known aberrations are applied. For more detailed information about the I-WFS, we direct the reader to the original papers\cite{ingot0,ingot2,ingot3,Carmelo2020} and further developments\cite{ingot4,ingot5,ingot6,ingot7,kalyan2020,Portaluri2020}.
\label{sec:intro}  

\section{The Ingot Wavefront Sensor}
\begin{figure}[h!]
\centering
\includegraphics[width=0.9\linewidth]{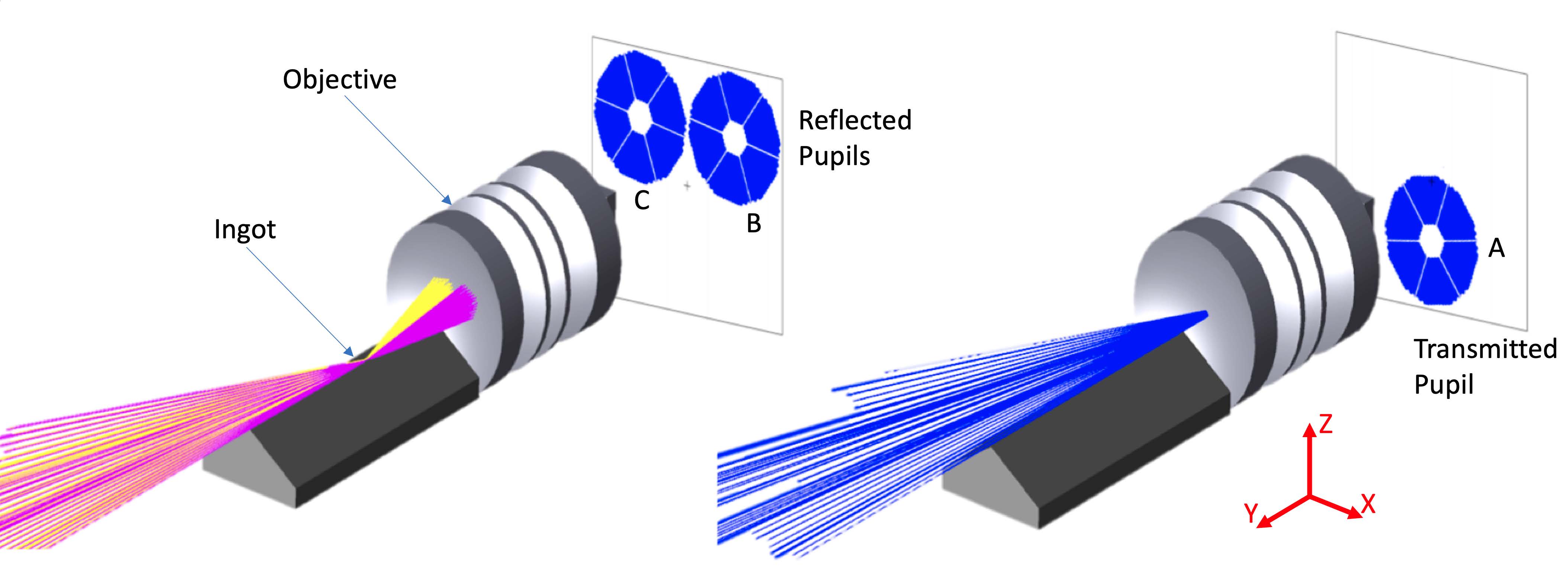}
\caption{Conceptual layout of the ingot prism. Part of the LGS is focused by the reflecting ingot roof and forms two pupils while the remaining part of the LGS is focused after the ingot and is transmitted directly to the pupil re-imaging optics forming the third pupil}
\label{fig:triangolo}
\end{figure}

According to the current design the I-FWS is composed of a reflective prism, which allows the splitting of the incoming LGS light onto three beams. Two of them are reflected and the third is passing unperturbed toward the following objective. These three beams are then re-imaged in three pupils at the conjugated plane as shown in Figure \ref{fig:triangolo}.

\begin{figure}[h!]
\centering
\includegraphics[width=0.6\linewidth]{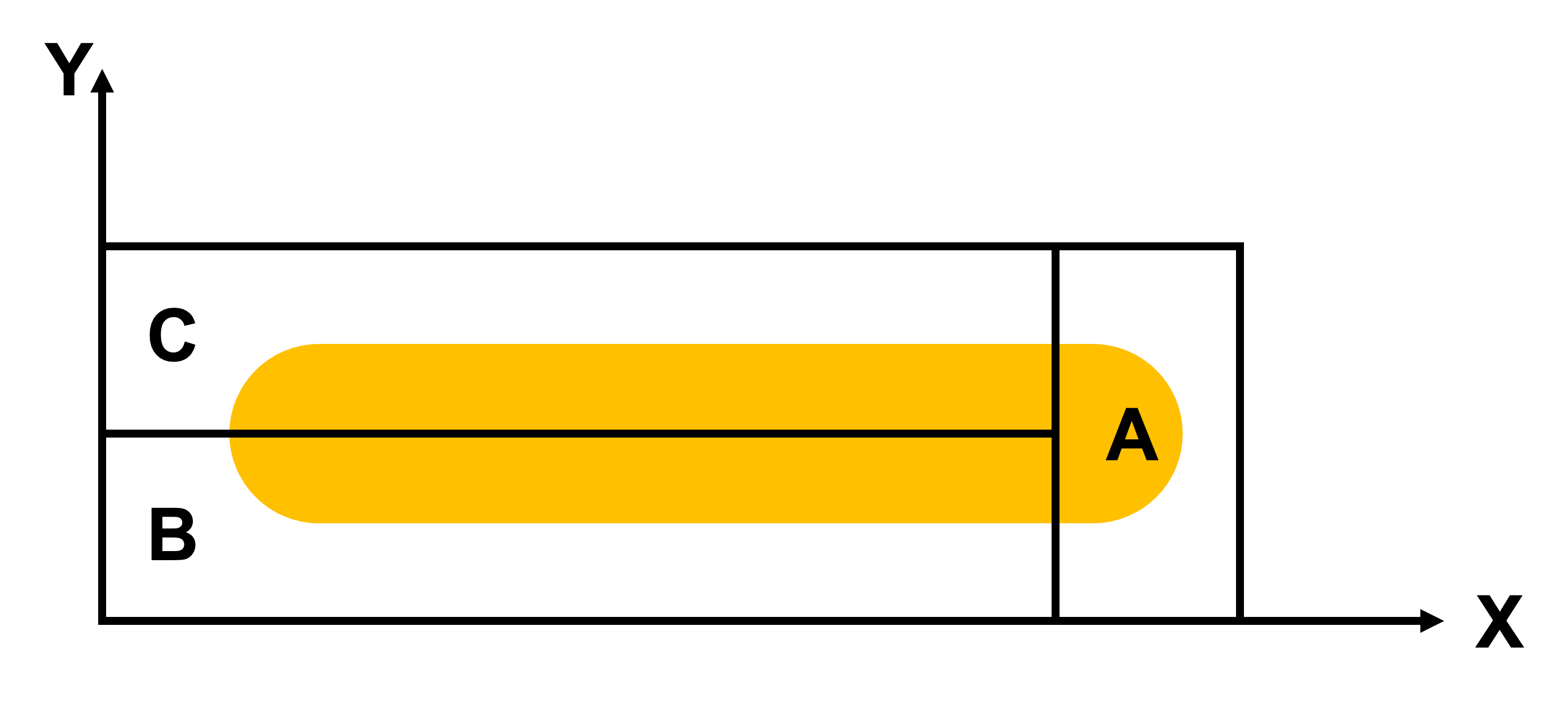}
\caption{}
\label{fig:3segnali}
\end{figure}

In agreement with the four faces Pyramid Wavefront Sensor\cite{pyramid} theory, the pupils are used to calculate signals proportional to the first derivative of the wavefront. In the following equation, we define the signals $S_x$ and $S_y$ along the x and y directions (x denotes movements orthogonal to the elongation direction, while y denotes movements along the elongation direction), according with Figure \ref{fig:3segnali} as:
\begin{align}
\label{eq1}
\begin{split}
 S_x = \frac{B-C}{A+B+C}-\frac{B_{ref}-C_{ref}}{A_{ref}+B_{ref}+C_{ref}} ,
\\
 S_y = \frac{A}{A+B+C}-\frac{A_{ref}}{A_{ref}+B_{ref}+C_{ref}}.
\end{split}
\end{align}
The operation is done pixel-wise such that a signal $S_x$ and $S_y$ is obtained for every pixel. Note that we subtracted the reference signal (also called reference slope), which is the signal that we consider the zero point in our calibration.

\section{The Ingot-WFS test bench: method and data analysis}

To investigate the behavior of the I-WFS not only from a simulation point of view, at the Observatory of Padua - INAF, we realized a test bench that aims to reproduce, as close as possible, the ELT main geometrical characteristics. We used simulations and laboratory data to compare, learn, and define a robust, entirely automatized alignment procedure using the optical feedback provided by the I-WFS.

\begin{figure}[h!]
\centering
\includegraphics[width=0.9\linewidth]{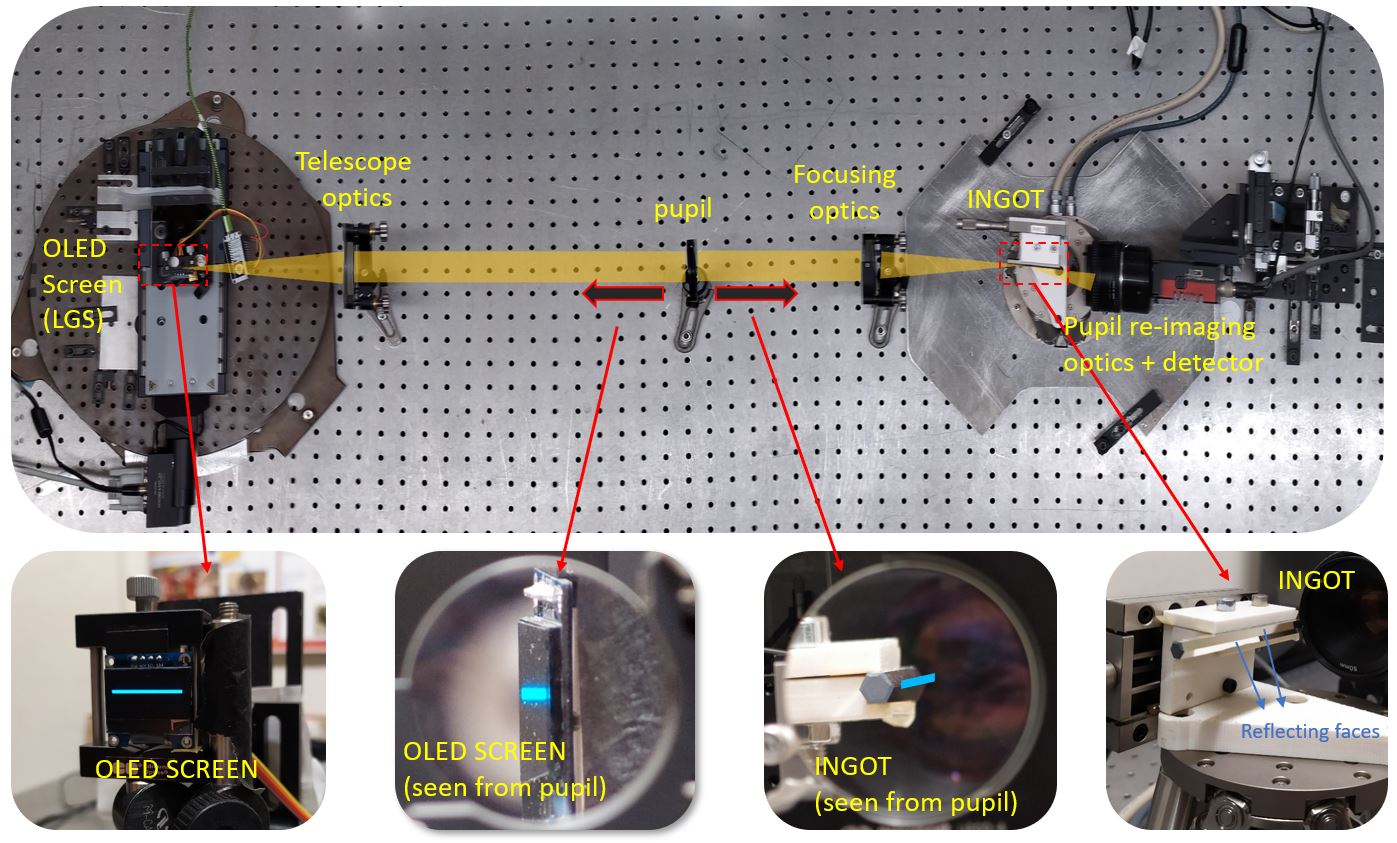}
\caption{Top view of the optical bench setup used to test the ingot alignment procedure. }
\label{fig:bench_ingot}
\end{figure}
\begin{figure}[h!]
\centering
\includegraphics[width=0.9\linewidth]{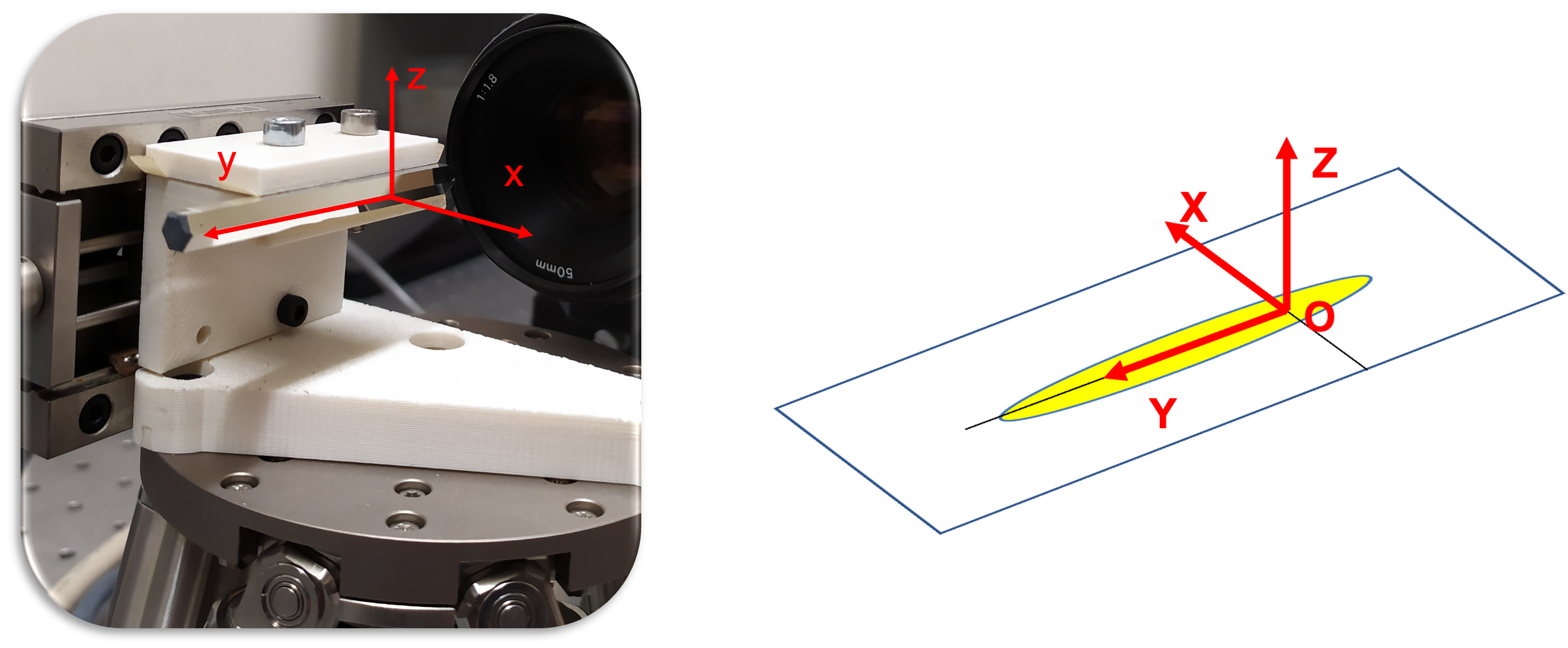}
\caption{Coordinate system of the Ingot prism}
\label{fig:sistema_rif}
\end{figure}

The test-bench setup is shown in Figure \ref{fig:bench_ingot}, is equipped with off-the-shelf components. It is specifically designed to simulate the imaging of the LGS source into the I-WFS prism and to produce an image of the three pupils on a dedicated camera with the proper sampling. 
The design aims to be as much as possible similar to the ELT geometry including:
\begin{itemize}
    \item {The ratio between the separation of the Laser Launcher Telescope from the optical axis and the entrance pupil diameter}
    \item{The physical size of the image on the I-WFS prism}
    \item{The telecentric image space (on the I-WFS prism)}
    \item{The aspect ratio of the LGS (spot elongation to FWHM ratio)}
\end{itemize}
The optical layout is a 1:1 re-imaging relay, composed of:
\begin{itemize}
    \item {An SSD1306 monochrome OLED display, with a format of 128x64 pixels with 170 $\mu m$ pixel size to reproduce the LGS. The advantage of this a device is that it has almost no background noise emission of black pixels, which is commonly present in other display technologies.}
    \item{An achromatic doublet with focal length f=200 mm that collimates the light from the LGS source, and another one that refocuses the light onto the I-WFS prism.}
    \item{A diaphragm located between the lenses that acts as the aperture stop/pupil of the system with a clear aperture of 25 mm. It is placed such that the system is telecentric in the image space, i.e. the stop is at the focus of the camera doublet.}
    \item{A pupil re-imager optics, consisting of a f=50 mm wide-aperture photographic objective.}
    \item{The I-WFS based on an hexagonal light pipe, generally used for beam homogenisation, whose external faces have been aluminized to obtain a reflective roof with an apex angle of $120^{\circ}$. Notice that this angle is the same one required at the ELT to achieve the proper separation of the reflected pupils.}
    \item{The camera, a Prosilica GT3300 from Allied Vision. The detector is an 8 Megapixel CCD sensor with 5.5 $\mu m$ pixel size. We use a 4x4 binning to sample the pupil diameter with $\approx 130$ pxl.}
\end{itemize}
The I-WFS prism is mounted on a H-811 Hexapod from Physik Instrumente. This device allows precision movements in all six degrees of freedom of the I-WFS where its position is defined by the reference coordinate system shown in Figure \ref{fig:sistema_rif} that is given by a rotation of $-94^\circ$ on the X-axis and a shift of -14 mm on the Y-axis with respect to the light path reference system.
 \\
The origin of the coordinates system is placed at the beginning of the Ingot prism facing the camera; the Y-axis coincides with the edge between the reflective faces of the Ingot with positive pointing towards the telescope. The Z-axis points in the vertical direction with respect to the bench. The X-axis is chosen to form a right-handed coordinate system. Finally, the pivot point for rotations is set at the origin of the coordinates system. Using the ray-tracing software Zemax-OpticStudio, we performed several simulations to investigate the response of the I-WFS to the six degrees of freedom misalignments. 
\newpage

We developed a fully automatic python procedure that, starting from the pupil frame is able to perform autonomously the following actions:

\begin{itemize}
    \item {Using a Canny-edge algorithm \cite{cannysigma}, the code is able to identify the 3 pupils position, in order to fit them and define the radii at the coordinates of the centers (in pixels), as shown in Figure \ref{fig:detection}
    \begin{figure}[h!]
    \centering
    \includegraphics[width=0.85\linewidth]{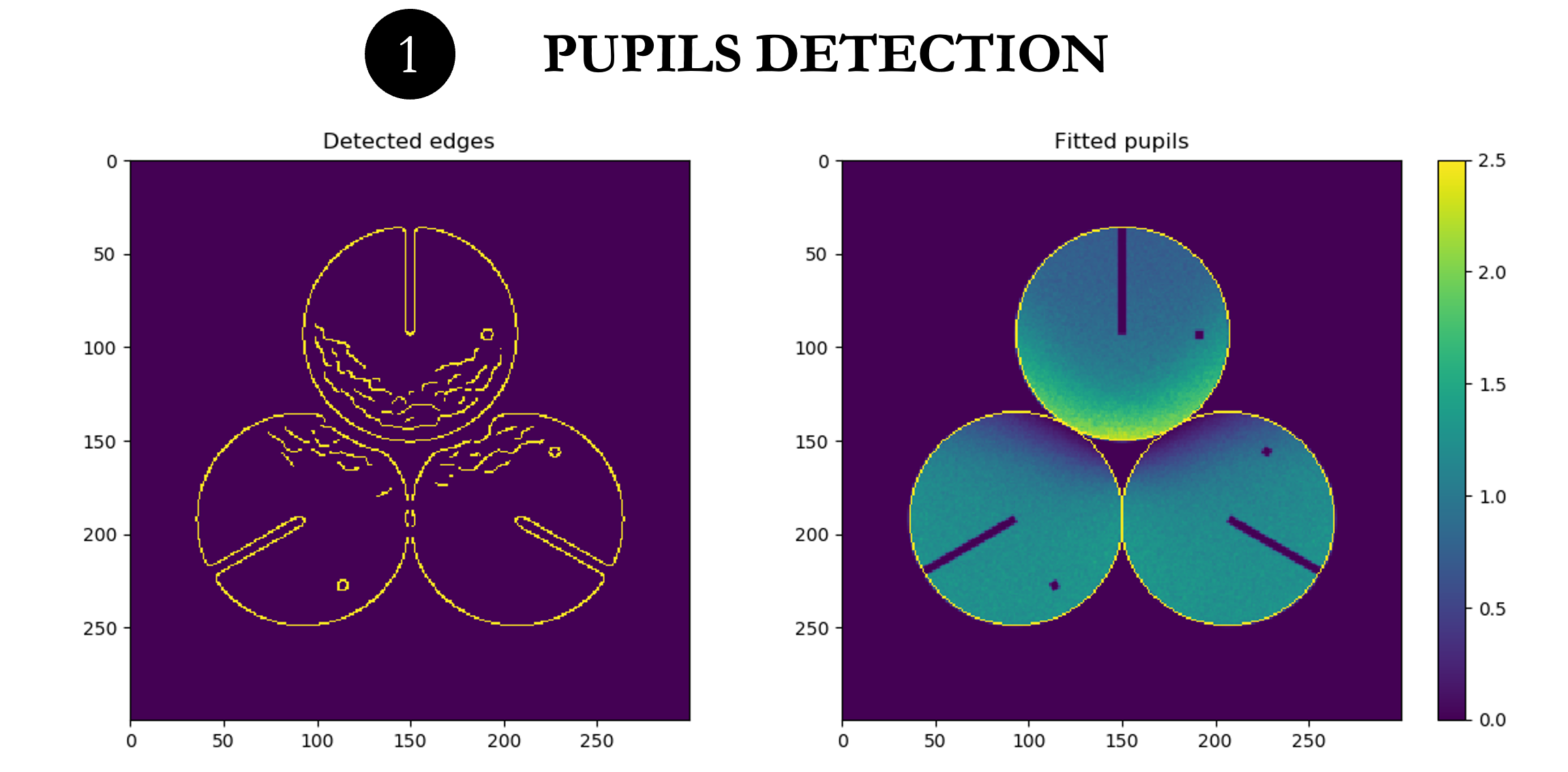}
    \caption{Left: Detection of the three pupils using the Canny-edge algorithm. Right: Best fit of the circular external edge}
    \label{fig:detection}
    \end{figure}}
    \item{Extract the three pupils from the image, perform a rotation and flip to compensate the effects induced by the reflection from the Ingot prism, as shown in Figure \ref{fig:extraction}
    
    \begin{figure}[h!]
    \centering
    \includegraphics[width=0.9\linewidth]{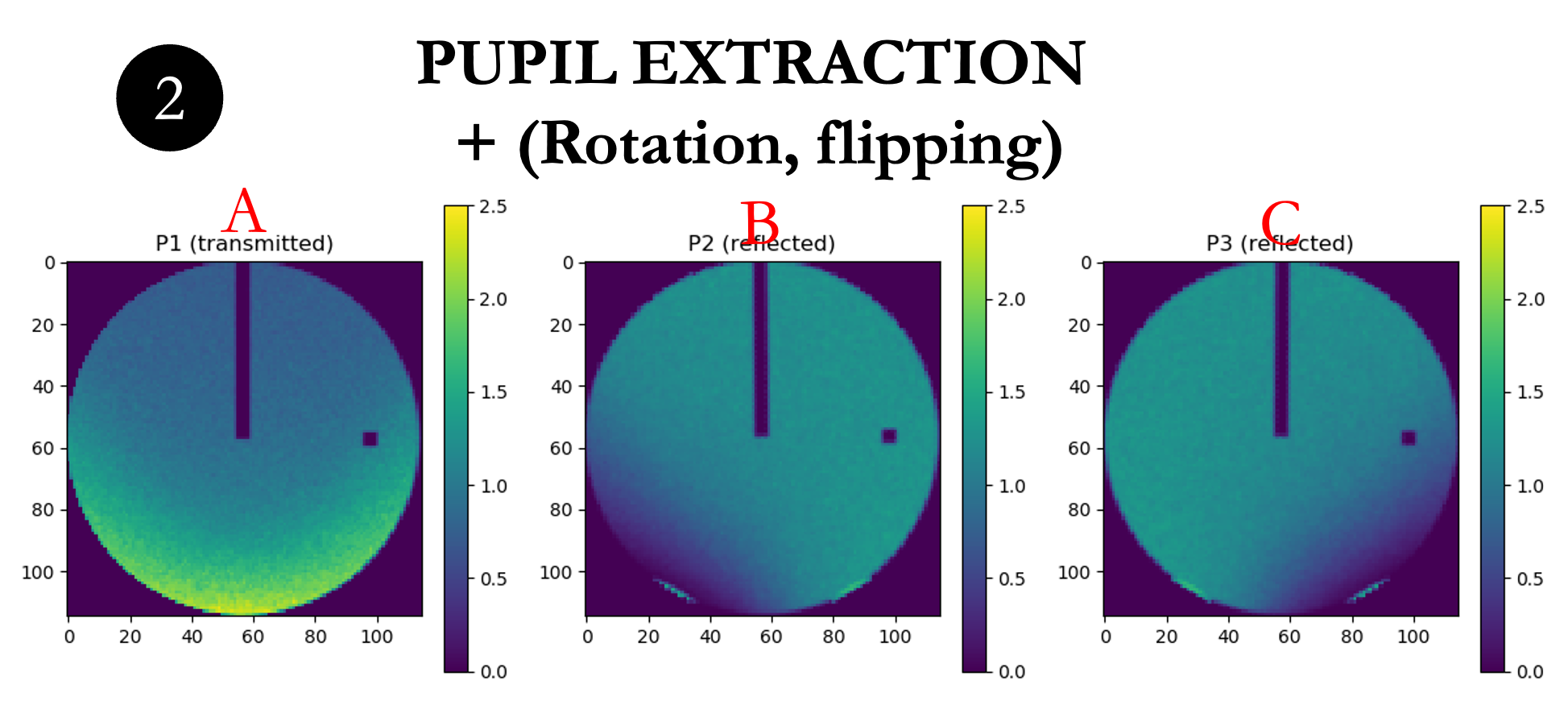}
    \caption{Left: Extraction of the pupils. Rotation and Flipping for the two reflected pupils}
    \label{fig:extraction}
    \end{figure}}
    \vspace{20pt}
    \item{Once the pupils are shown and co-aligned, they are used to calculate the signals as indicated in equation \eqref{eq1} and the result obtained appear in Figure \ref{fig:signals}
    
    \begin{figure}[h!]
    \centering
    \includegraphics[width=0.7\linewidth]{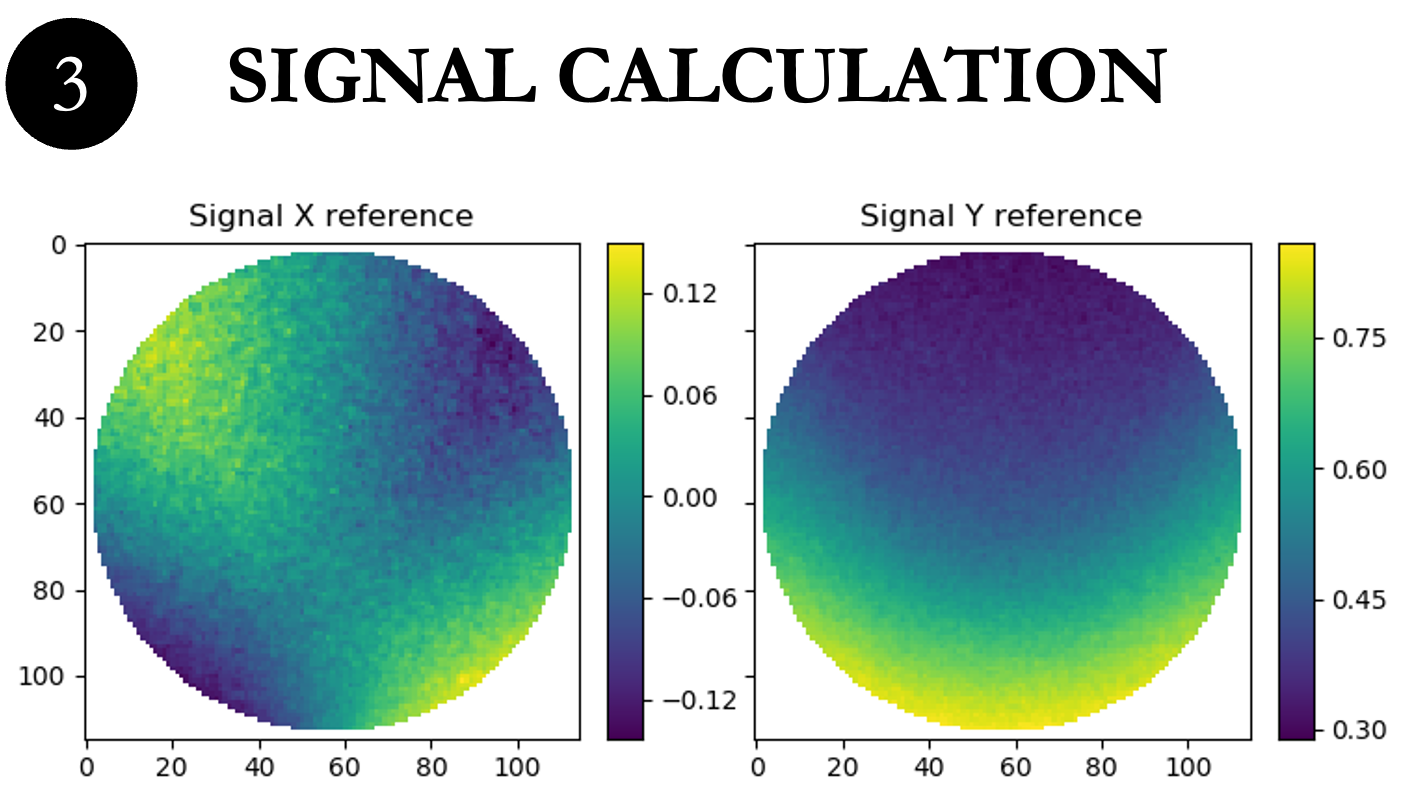}
    \caption{Signals $S_x$ and $S_y$}
    \label{fig:signals}
    \end{figure}}
\end{itemize}
Note that in Figures \ref{fig:detection} and \ref{fig:extraction}, we added a line and a small square to highlight the orientation and to show how pupils B-C are rotated and flipped due to the reflection from the Ingot prism.

At this point, we are able to investigate the effects of misalignment of the ingot prism with respect to the LGS. We performed a sensitivity analysis by moving the ingot along each degree of freedom (dof), one at a time,  and we identified the following observables as a gauge of the alignment of the I-WFS:
\begin{itemize}
    \item {Distance between the pupils}
    \item{Integrated flux within each pupil}
    \item{The variation produced on the signals $S_x$ and $S_y$}
\end{itemize}
Note that, in the aligned condition, the edge of the Ingot prism is placed along the axis of symmetry of the LGS image, the three pupils are equidistant from each other, and both reflected pupils have the same integrated flux. The amount of light in the transmitted pupil is still a free parameter, that for convenience we set equal to the light on the reflected pupils, to have $1/3^{rd}$ of the total flux in each pupil, thereby maximizing the dynamic range available on the detector. The outcome of the analysis, is summarized in the table \ref{tab:dioporco} and fully described in the previous work \cite{kalyan2020}. In the table, we distinguish between reflected and transmitted pupils and to give an idea of the magnitude of the change we use (+ +) for stronger variations and (+) for weaker variations.
\newline
\begin{table}[h!]
\centering                       
\begin{tabular}{|l|l|l|l|l|}   
\hline               
\textbf{Degree of Freedom} & \textbf{A Flux} & \textbf{B \& C Flux} & \begin{tabular}[c]{@{}c@{}}\textbf{Separation}\\ \textbf{A-B or A-C}\end{tabular} & \textbf{Separation B-C} \\  
\hline
    Decenter X  & +  & ++ &    &    \\
    Decenter Y  & +  & +  &    &    \\
    Decenter Z  & ++ & ++ &    &    \\
    Tilt around X &    &    & ++ & ++ \\
    Tilt around Y &    &    & ++ &    \\
    Tilt around Z &    & ++ & ++ & +  \\
\hline   
\end{tabular}
\caption{Degrees of freedom and the corresponding observables}
\label{tab:dioporco} 
\end{table}
\vspace{50pt}
\newpage

The analysis performed above is the key to build a powerful alignment procedure of the I-WFS with respect to the LGS source. We developed a procedure that is composed of two levels: the first is related only to the position of the pupils and the distribution of light (in the sense of integrated flux within each pupil). The aim here is to align the I-WFS to the optical axis of the telescope and the fore-optics. The second level takes into account the movements of the LGS on sky, which, from a practical point of view, represent a movement of its image on the I-WFS, producing a variation of the flux and signals. Note that movement of the source on-sky does not produce a change in the position of the pupils, because the latter depends solely on the relative alignment between the ingot and telescope. Thus, from an operational point of view, the first alignment is related to the alignment of the ingot WFS to the telescope (first alignment and any other re-alignment necessary to compensate flexures or other movements of the optics), while the second alignment is related to the compensation of LGS position (focus variation or jitters of the LGS). The procedure and tests presented below are ideal because they do not consider any mid-high order aberration. After successfully identifying the starting position, we developed a fully automatic alignment procedure that aims to speed up the tests and to perform a statistical analysis of the convergence. This has been done by remotely controlling the Hexapod and the camera, through a Python script that was specifically written to automatize the calculation of pupils positions, light distribution and signals, in order to allow reconstruction and compensation processes.

To perform the procedure, we have considered 6 observables, that are linearly independent, and listed below:

\begin{itemize}
    \item {1. The difference between the flux in the (B-C) pupils, normalized to the total flux of the pupils: left panel of Figure \ref{fig:duedi}.}
    \item {2. The flux of the A-(B+C) pupils, normalized to the total flux of the pupils: right panel of Figure \ref{fig:duedi}.
    \hspace{10pt}

    \begin{figure}[h!]
    \centering
    {{\includegraphics[width=8cm]{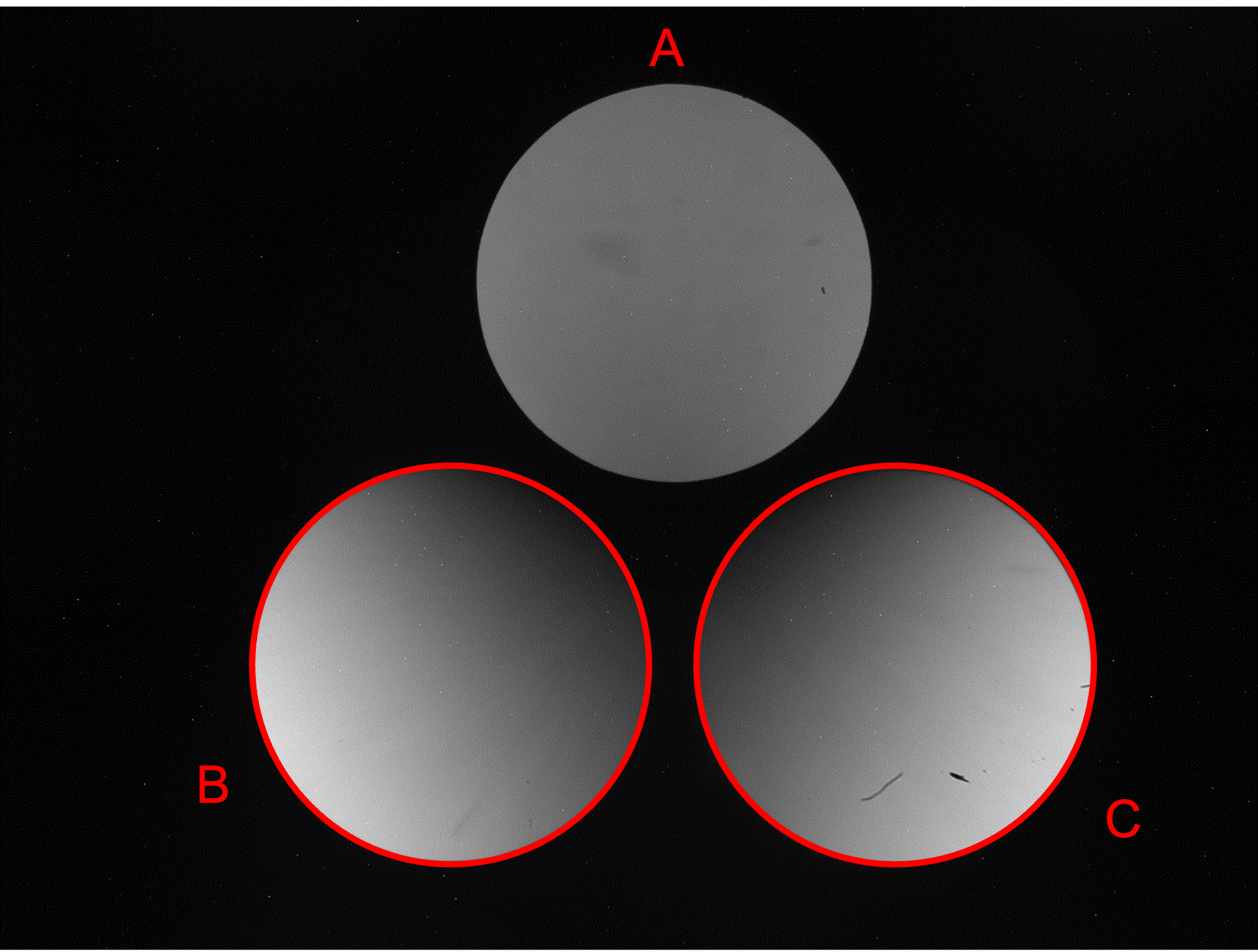}}}
    \qquad
    {{\includegraphics[width=8cm]{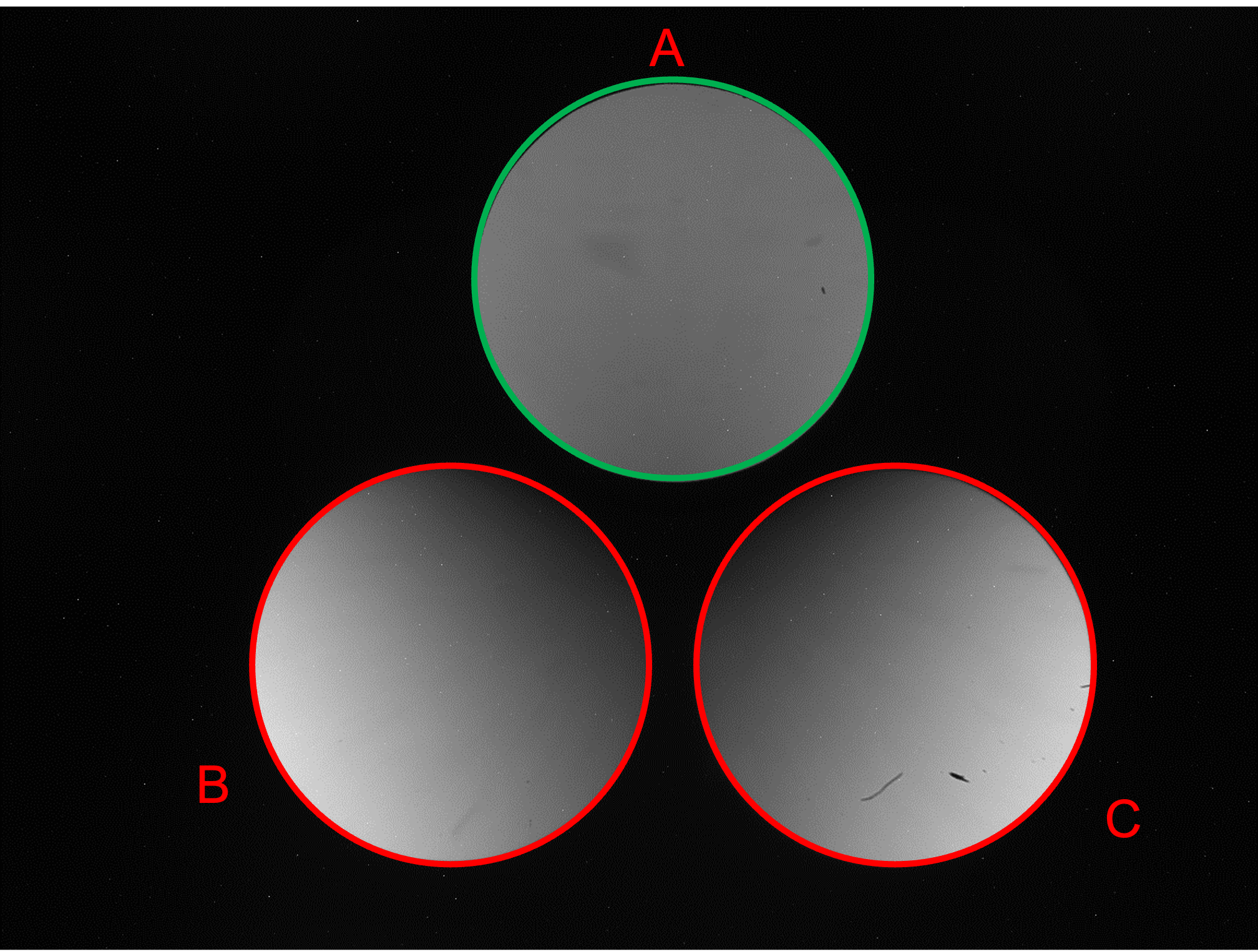}}}
    \caption{Graphical representation of the observables 1 (left) and 2 (right).}
    \label{fig:duedi}
    \end{figure}}

\end{itemize}
\newpage
\begin{itemize}
    \item {3. The separation between the centers of the A \& B pupils: left panel of Figure \ref{fig:tredi}.}
    \item {4. The separation between the centers of the A \& C pupils: right panel of Figure \ref{fig:tredi}.

    \begin{figure}[h!]
    \centering
    {{\includegraphics[width=8cm]{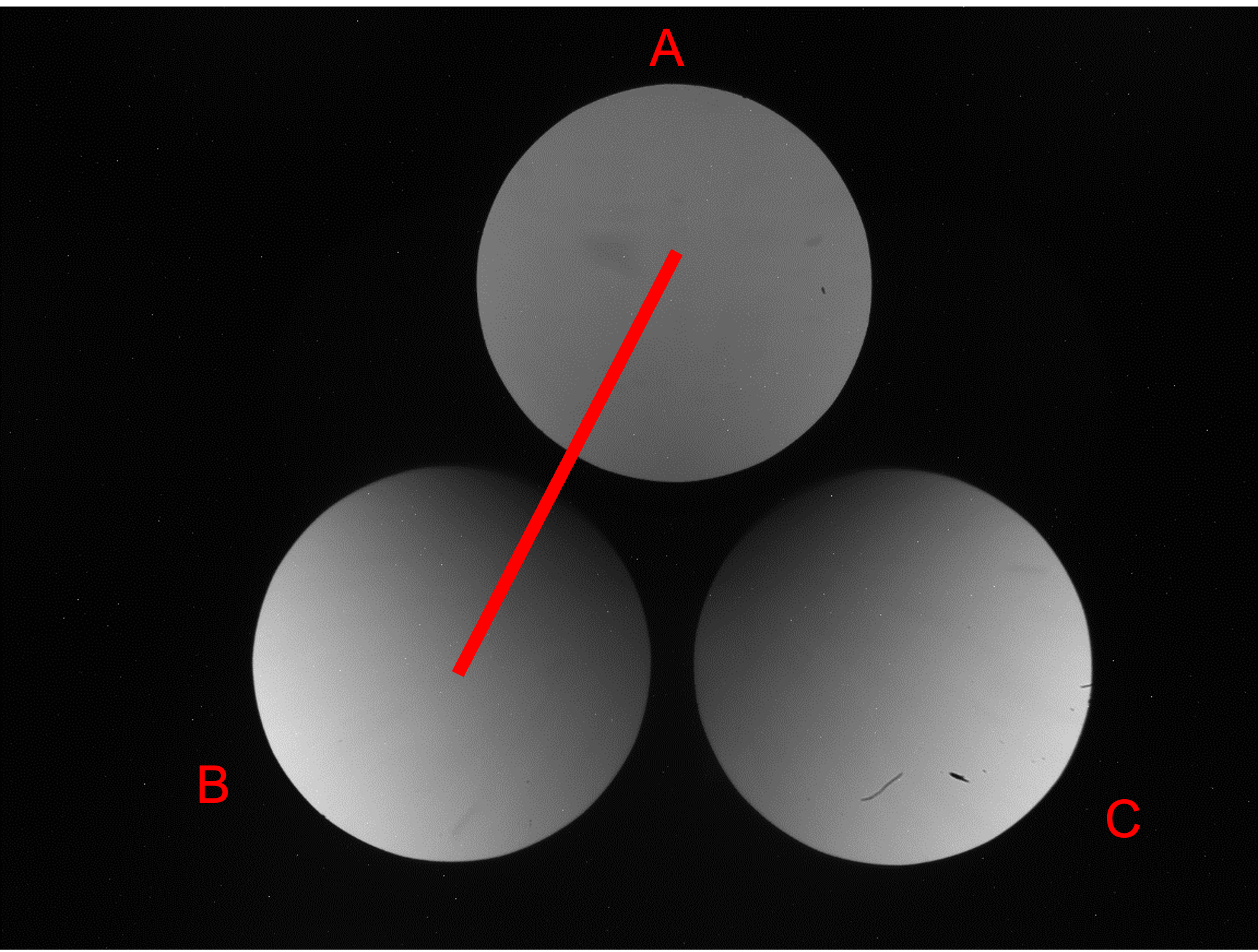}}}
    \qquad
    {{\includegraphics[width=8cm]{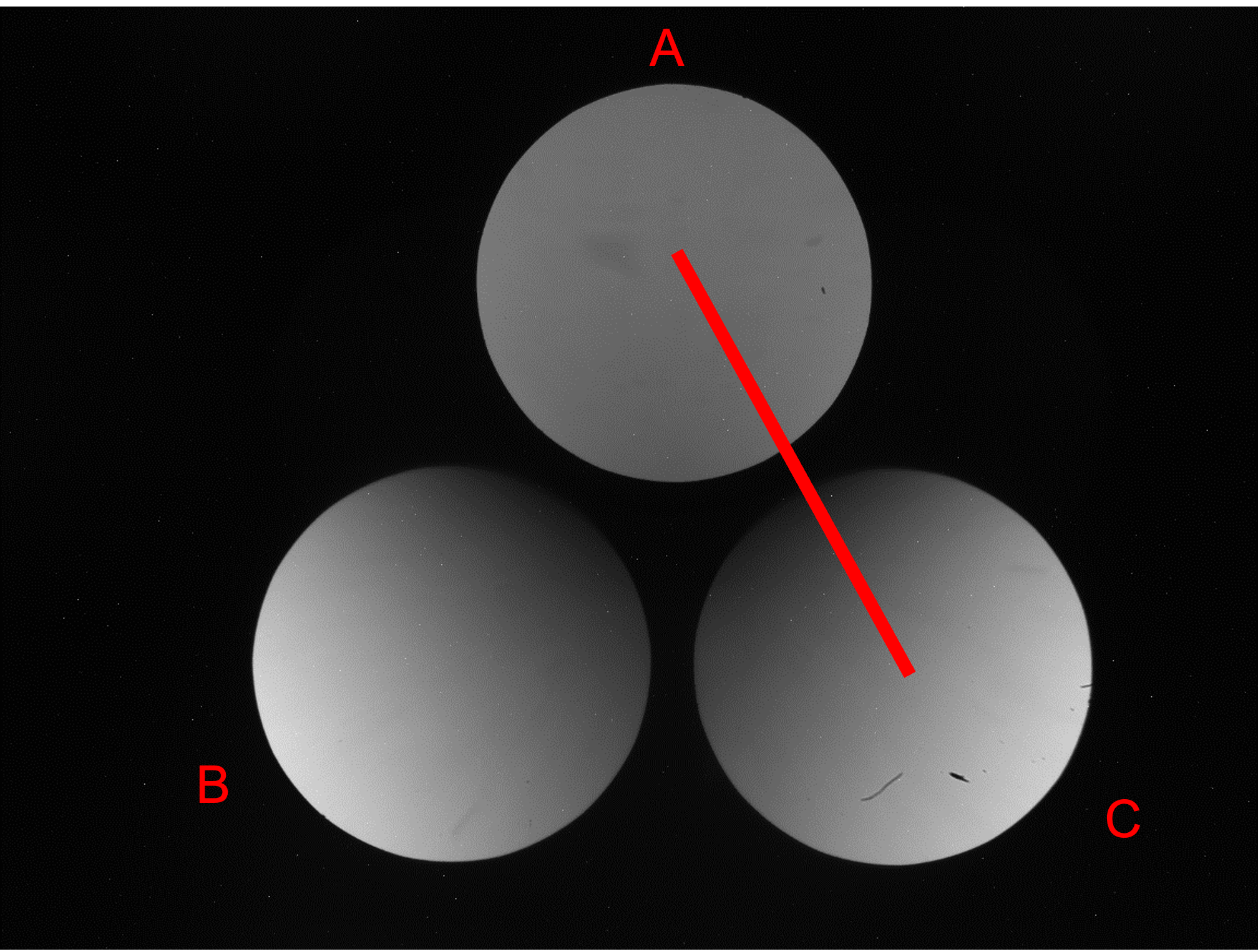}}}
    \caption{Graphical representation of the observables 3 (left) and 4 (right).}
    \label{fig:tredi}
    \end{figure}}
\end{itemize}
\begin{itemize}    
    \item{5. The position of the centers of the B \& C pupils with respect to the Y-axis: left panel of Figure \ref{fig:4di}}
    \item{6. The difference of the average of $S_x$ in two regions that correspond to the top-left (1) and top-right (2): right panel of Figure \ref{fig:4di}. Note that the size is chosen to not reach the middle of the pupil.
    \begin{figure}[h!]
    \centering
    \includegraphics[width=8cm]{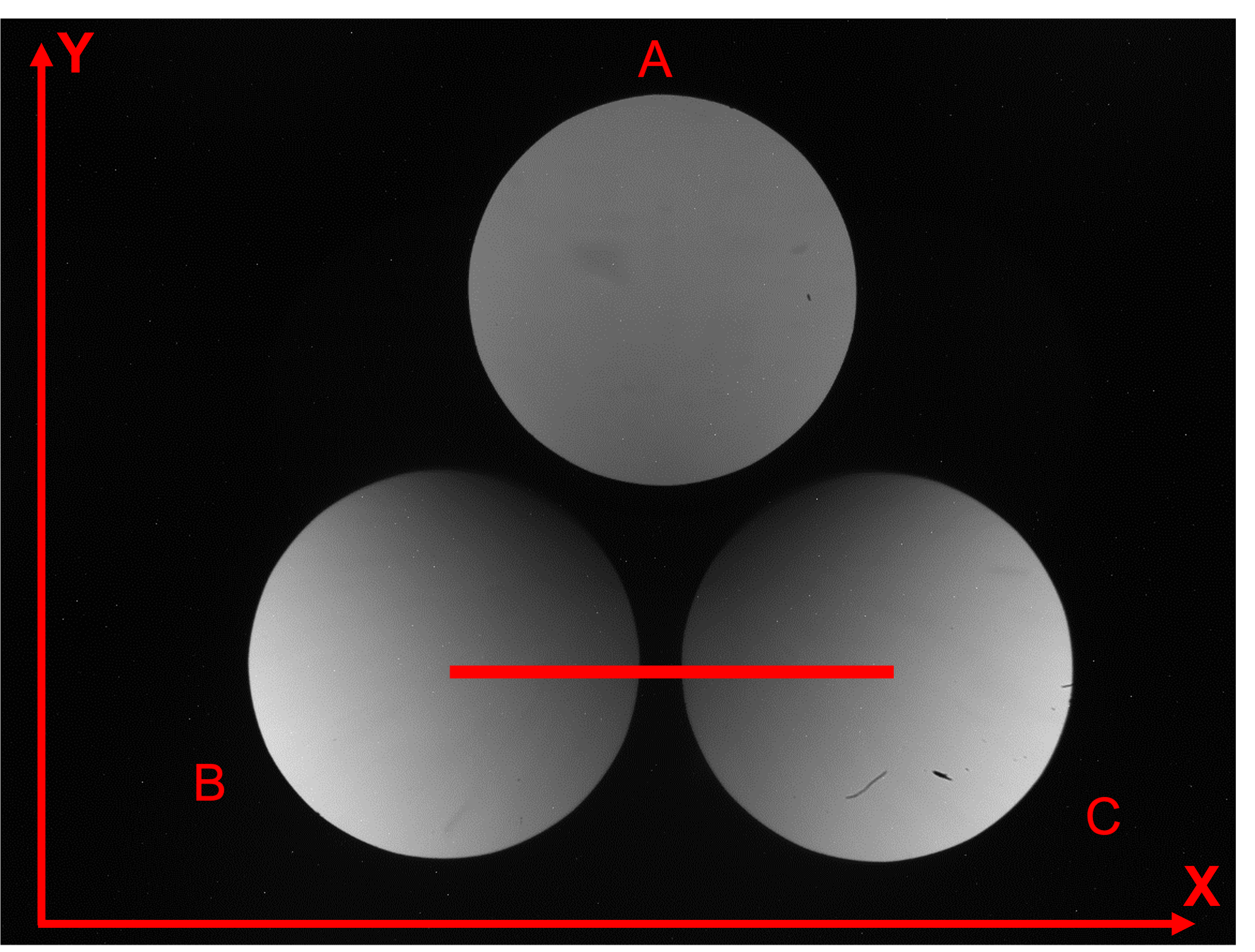}
    \qquad
    \includegraphics[width=7cm]{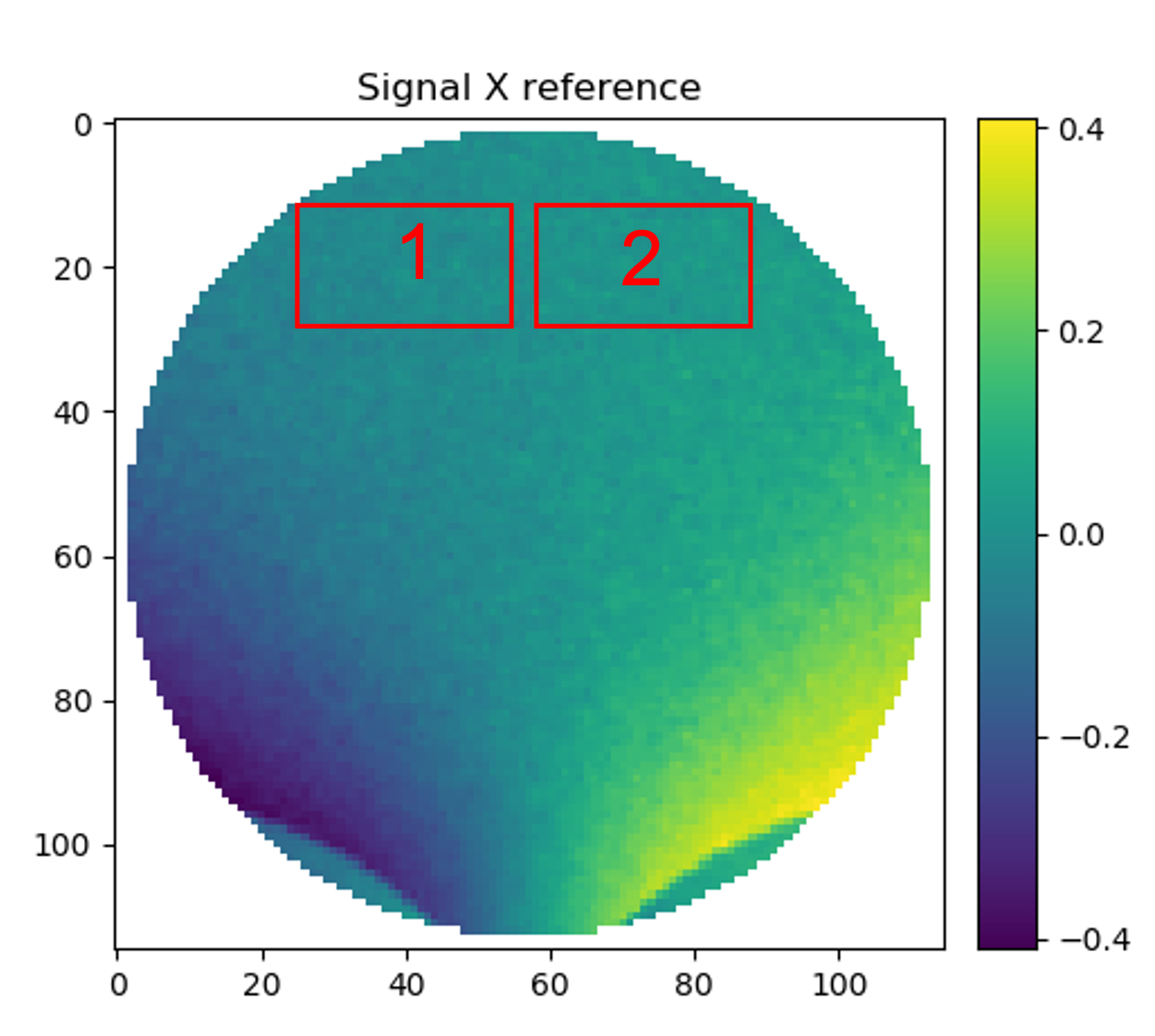}
    \caption{Graphical representation of the observables 5 (left) and 6 (right).}
    \label{fig:4di}
    \end{figure}}
\end{itemize}

In order to let our system know that the alignment position is reached, we have defined a target value for each observable listed above. Table \ref{tab:observables} lists the observables and target values.
\begin{table}[h!]
\centering                       
\begin{tabular}{|c|c|}
\hline
\textbf{Observables}                       & \textbf{Target Values} \\ \hline
\hline
Flux (B-C)/Flux (A+B+C)     & 0          \\
\hline
Flux (B+C-A)/Flux (A+B+C)   & 2/3        \\
\hline
Distance A-B                & 277 pxl    \\
\hline
Distance A-C                & 277 pxl    \\
\hline
Position B-C                & 0          \\
\hline
Avg Sx{[}1{]}-Avg Sx{[}2{]} & 0          \\ \hline
\end{tabular}
\caption{Observables used to build the alignment procedure, together with the target values that our system should ideally reach}
\label{tab:observables}     
\end{table}

\begin{figure}[h!]
\centering
\includegraphics[width=0.4\linewidth]{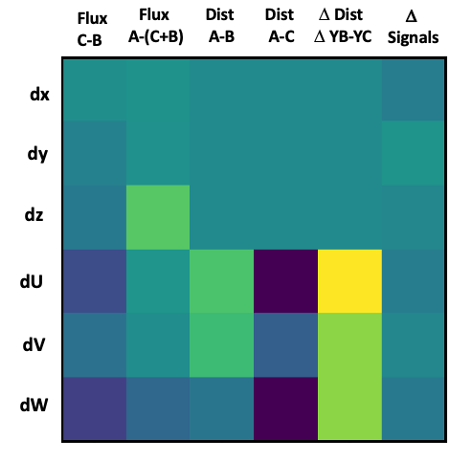}
\caption{Interaction Matrix description}
\label{fig:IM}
\end{figure}

\begin{figure}[h!]
\centering
\includegraphics[width=0.9\linewidth]{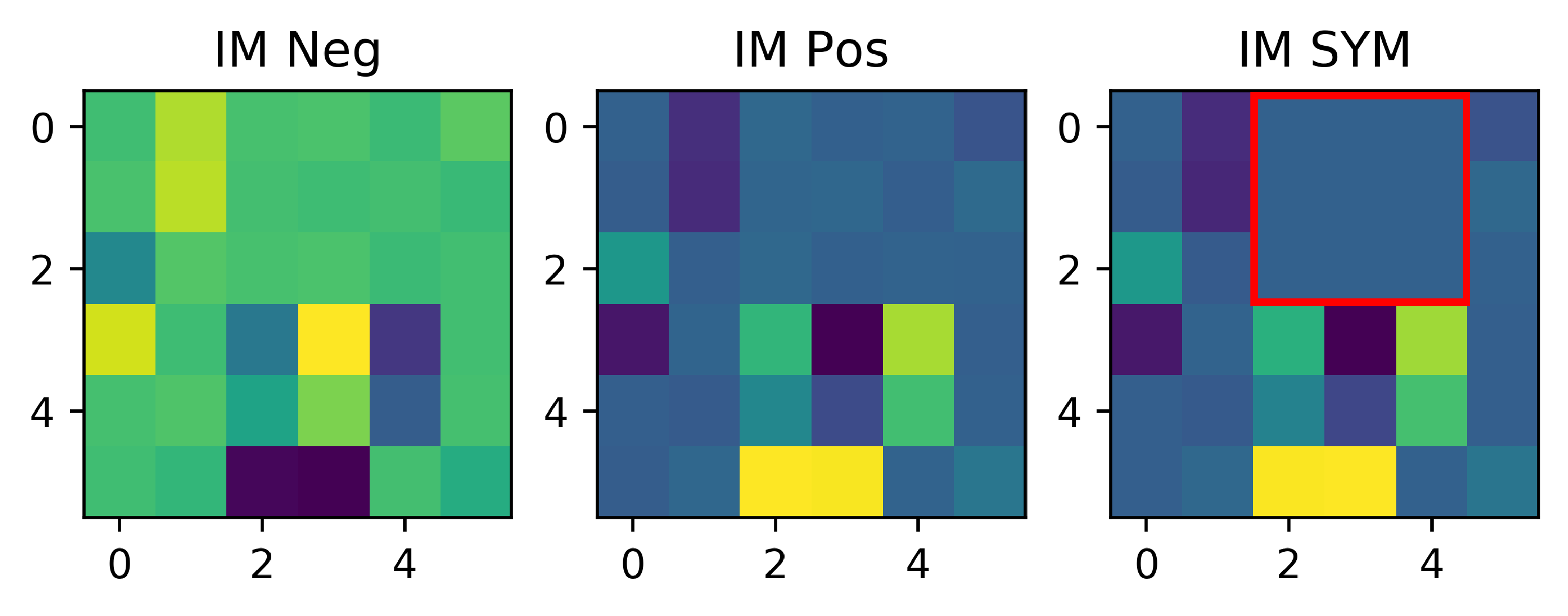}
\caption{Interaction Matrices. Left: IM computed on pull. Center: IM computed for push. Right: IM computed as a combination of push and pull. The values in the red box have been fixed to zero because they are not affected by the decenter along the three axis. Note that on X axis are shown the observables, while on Y axis are shown the degrees of freedom of the system.}
\label{fig:IMs}
\end{figure}

The alignment procedure is based on a linear approximation approach similar to that commonly used in AO closed-loop theory. We calculated an Interaction and Control matrix (IM $\&$ CM), as is usually done for Deformable Mirrors (DM). In our context, IM and CM do not take into account the Zernike modes of the DM, but rather the variation of the 6 observables due to misalignments of the Ingot prism with respect to the optical axis of the system. The calculation of the IM and CM is also automatized through a dedicated Python routine, which uses the first rough alignment as a starting point for the calibration of the interaction matrix.
A detailed description of the calibration procedure is given in the following:
\begin{itemize}
    \item {1. Start from an hexapod (i.e. Ingot prism) position that should be close to the position that we have identified as the aligned one. This makes the procedure faster from a computational point of view.}
    \item{2. Perform the computation of the 6 observables defined in table \ref{tab:observables}.}
    \item{3. Define a set of 6 coordinates to be used as misalignments to displace the hexapod stage as is done with DM. We apply each dof at the time, as push and pull manner, performing each time the observables computation. After this process, we expect to have for each dof applied 2 sets of 6 observables: One for the push (positive dof) and one for the pull (negative dof).}
    \item{4. Calculate the difference for each dof between the observables computed in step 3, with respect to that computed in step 2.}
    \item{5. Build the Interaction Matrix starting from the results of point 4. Obtaining 2 IM: one coming from the push, and one from the pull. These two IMs will be averaged into a single one. Thereafter, it will be pseudo-inverted in order to produce the Control Matrix, as shown in Figure \ref{fig:IMs}.}
\end{itemize}
In order to have a more clear understanding of how the IMs is structured, we report an example in Figure \ref{fig:IM}, where the rows contain the 6 dof, while the columns hold the 6 observables.
In Figure \ref{fig:IMs}, are shown the IMs produced for the push/pull and one that is the combination of them. Note that, to avoid propagation of noise, we fixed to zero the terms in the red box of the so-called \textit{IM SYM}. This is because these values are not affected by the decenters but only by tilts. 
\newline
Once the calibration is performed, we use the CM (i.e. RM) to reconstruct the misalignments of the system with respect to the target position, that we want to reach as the goal of the alignment process. To validate the calibration, we performed the steps listed below:
\newline
\begin{itemize}
    \item {Move the Hexapod to a random misaligned position.}
    \item {Compute the difference between the observables calculated in this position and the observables target values shown in table \ref{tab:observables}}
    \item{Multiply this difference by the RM, obtaining the corrections}
    \item{Multiply these corrections by the amplitude of each dof that we used in the step 3 of the calibration process}
\end{itemize}

Once the Ingot prism is properly aligned to the system, it is time to pay attention on the movements of the LGS in the sky. This corresponds to the movement of the LGS image on the Ingot prism. This produces a variation of the flux and signals, but, once again, it will not change the position of the pupils. Thus, the distance between them cannot be used to keep the LGS image aligned to the Ingot prism during operations. The procedure is based solely on the measurements of signals (only the low order modes) to keep the alignment between Ingot and source. In the lab, the misalignment between Ingot and source is produced by moving the Ingot via the hexapod and not by moving the source with respect to the telescope simulator. This is because, according to the current design of the bench, the source position cannot be controlled precisely through a hexapod, but with a good approximation, the effect produced on the signals by the movement of the hexapod is equivalent to that produced by a movement of the source, because of the symmetry between object and image space. As before, the strategy adopted to reconstruct misalignments between the Ingot and the LGS image is based on a linear approximation (interaction matrix calculation), and fully described in the previous work \cite{kalyan2020}. 
\section{The deformable lens}
\begin{figure}[h!]
\centering
\includegraphics[width=1.0\linewidth]{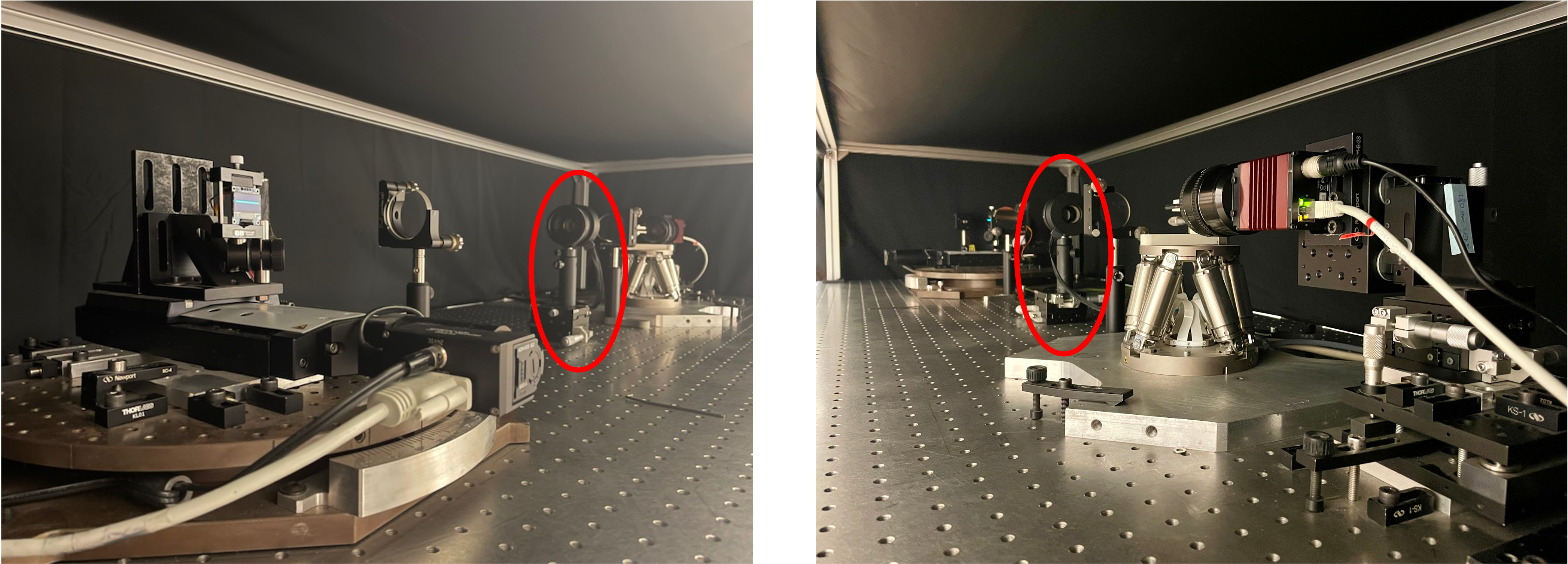}
\caption{View of the I-WFS test bench, equipped with a Deformable Lens conjugated to the pupil plane.}
\label{fig:doppia}
\end{figure}

We equipped the bench with a Deformable Lens (DL)\cite{Bonora} positioned in the pupil plane, as shown in Figure \ref{fig:doppia}. This device, that is able to introduce low order known aberration terms to the wavefront, and presents two notable advantages compared to other possible solutions: does it not require to modify the existent optical path and its installation is extremely straightforward, common off-the-shelf lab parts can be used to hold it. The DL is the AOL1825 developed by the Dynamic Optics, with a clear aperture of 25.5mm and 18 actuators (9 on each side of the lens). 

\begin{figure}[h!]
\centering
\includegraphics[width=1\linewidth]{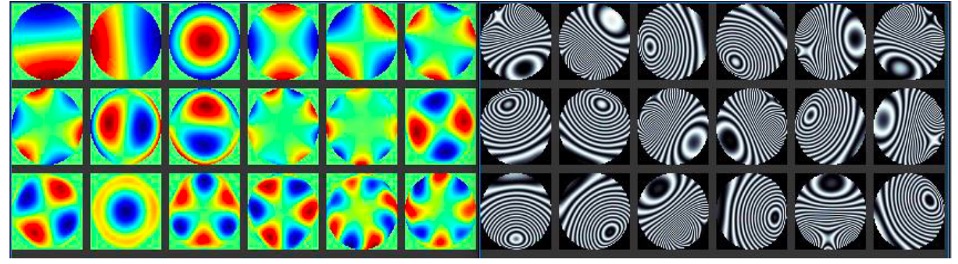}
\caption{Left: Aberration modes produced by the 18 actuators of the DL. Right: Corresponding actuators shape}
\label{fig:modes}
\end{figure}

The DL is able to apply the first 18 Zernike modes as shown in Figure \ref{fig:modes}, and can be controlled both using its own software or the Matlab custom libraries developed by the owner. 
\newpage
In this framework, we developed a Python/Matlab procedure to control the system. We are now able to apply a single value or a combination of modes and actuators. We also defined a calibration procedure, commonly used in AO closed-loop theory, that aims to calculate an Interaction and Control matrices. The procedure is composed as follow:
\newline
\begin{itemize}
    \item {1. Select the first 9 modes}
    \item {2. Collect frames for each mode in push/pull manner for 20 times}
    \item {3. Calculate the average of the frames and the signals $S_X$ and $S_Y$}
    \item {4. Average the $S_X$ and $S_Y$ for push and pull}
    \item {5. Make the fit for the 21 Zernikes modes to $S_X$ + the 21 Zernikes modes to $S_Y$}
    \item {6. Build the Interaction Matrix with a dimension of 42x9 and the Control Matrix via the SVD theory}
\end{itemize}
In addition, we added to the ray-tracing simulator an aberration surface, to have a model to compare and validate the laboratory results. We plan to investigate and fully characterize the sensitivity and linearity of the I-WFS when known aberrations are applied.
\section{Conclusion and Future}
The Ingot Wavefront Sensor has been proposed as an unconventional pupil plane wavefront sensor to cope with the natural elongation of the Sodium LGSs. In this work, we review the status of the laboratory testing of the I-WFS at the laboratory of the Astronomical Observatory of Padua, where we designed and developed a test bench that aims to operate in an open loop scenario. We presented a fully automatic Python procedure that aims to align the wavefront sensor both to the telescope and to the LGS movements in the sky. In addition, we reported the latest upgrade made with the Deformable Lens, which aims to investigate and produce a full characterization of the sensitivity of the I-WFS when known low-order aberrations are applied. 

We are planning to equip our test bench with a deformable mirror for the close-loops operations and eventually with a "more classical" wavefront sensor (Pyramid or Shack-Hartmann) to compare the performances. Last but not least, we are planning to participate to next year call at ESO Wendelstein Laser Guide Star Unit (WLGSU) @ William Herschel Telescope, where a sodium laser beam is launched as part of laser guide star field tests for validating ELT laser guide star baseline performance, as well described in the work \cite{Portaluri2022}.

\acknowledgments
We acknowledge the ADONI Laboratory for the support in the development of this project. Funding to support the laboratory activities described in this paper came also from the INAF Progetto Premiale ”Ottica Adattiva Made in Italy per i grandi telescopi del futuro”.


\end{document}